\documentclass[prb,twocolumn,superscriptaddress,showpacs,aps,letterpaper]{revtex4}

\usepackage{graphicx}

\newcommand{\MO}{$\eta$-Mo$_4$O$_{11}$}

\begin{document}

\title{Hidden one-dimensional electronic structure and non-Fermi liquid angle resolved photoemission line shapes of \MO}

\author{G.-H. Gweon$^*$}
\author{S.-K. Mo}
\author{J. W. Allen}

\affiliation{Randall Laboratory of Physics, University of Michigan, Ann Arbor, MI 48109}

\author{C. R. Ast$^\$$}
\author{H. H\"ochst}

\affiliation{Synchrotron Radiation Center, University of Wisconsin-Madison, Stoughton, WI 53589}

\author{J. L. Sarrao$^\dagger$}

\author{Z. Fisk$^\ddagger$}

\affiliation{National High Magnetic Field Laboratory, Florida State University, FL 32306}


\begin{abstract}
We report angle resolved photoemission (ARPES) spectra of \MO, a layered metal that undergoes two charge density wave (CDW) transitions at 109 K and 30 K\@.  We have directly observed the ``hidden one-dimensional (hidden-1d)'' Fermi surface and an anisotropic gap opening associated with the 109 K transition, in agreement with the band theoretical description of the CDW transition.  In addition, as in other hidden-1d materials such as NaMo$_6$O$_{17}$, the ARPES line shapes show certain anomalies, which we discuss in terms of non-Fermi liquid physics and possible roles of disorder.
\end{abstract}


\maketitle


\MO\ has a monoclinic crystal structure in which Mo$_6$O$_{22}$ layers, made up of distorted MoO$_6$ octahedra and defining a two-dimensional (2d) tetragonal plane ($bc$ plane), are connected weakly by MoO$_4$ tetrahedra \cite{Canadell93,Ghedira85}, giving rise to a quasi-two dimensional (quasi-2d) electronic structure \cite{Inoue88,Hill97}.  Within the octahedral layer, crossed conducting chains exist along $b$ and $b\pm c$ directions.  In this paper, we will refer to the three 1d bands defined on these chains as $b$ band and $b\pm c$ bands.  These 1d bands are nonetheless weakly hybridized and so \MO\ displays the ``hidden one dimensionality'' \cite{Canadell89,Whangbo91}.  The resulting Fermi surface (FS) \cite{Canadell89} can be viewed as being made up of pairs of parallel lines derived from each chain, with only slight distortions due to the weak hybridization between the chains.  The nominally 1d portions of the FS are ``nested,'' i.e.~separated by a constant wave vector \textbf{q}$_c$=2\textbf{k}$_{F}$, where \textbf{k}$_{F}$ is the FS wavevector, and so the term ``hidden nesting'' \cite{Whangbo91} was originally used to describe such materials.

One important aspect of the electronic structures of hidden 1d metals is that the nesting property can lead to the formation of a charge density wave (CDW) having the nesting wavevector \textbf{q}$_c$ \cite{Gruner94}.  The nested portion of the FS is then destroyed below the CDW transition temperature $T_c$ by the formation of gaps in the single particle electron structure.  \MO\ undergoes two successive CDW transitions  \cite{Guyot85} at $T_{c1}$ = 109 K and $T_{c2}$ = 30 K, with wavevectors \textbf{q}$_{c1} = 0.23$\textbf{b$^{*}$} and \textbf{q}$_{c2} = x$\textbf{a}$^{*}$ + 0.42\textbf{b}$^{*}$ + 0.28\textbf{c}$^{*}$ ($x$ is unknown), respectively, as determined by electron diffraction and X-ray diffuse scattering.  These wavevectors are in good agreement with calculated nesting vectors of the hidden 1d FS \cite{Canadell89,Hill97}.  A unique property of \MO\ is that it shows an anomalous bulk quantum Hall effect (QHE) at a very low temperature in the second CDW phase \cite{Hill98,Sasaki01}.  The FS topology is equally important in understanding the QHE, since the formation of the small cylindrical FS below $T_{c2}$, after most of the FS is destroyed by the imperfect nesting, is a central part of current the understanding of the QHE.

Another important aspect is that the underlying 1d character raises the possibility that the exotic non-Fermi liquid (NFL) properties of strictly 1d interacting electron models, e.g.~the electron fractionalization of the Luttinger liquid, may be present in one form or another in spite of the electron hopping between the chains.  In this respect, hidden 1d materials are significant \cite{Gweon03} for being intermediate between materials with weakly coupled parallel 1d chain, and fully 2d materials.

\MO~has interesting similarities and differences in the electronic structure, when compared to the well-known quasi-low-d ``purple bronze'' materials, Li$_{0.9}$Mo$_{6}$O$_{17}$, NaMo$_{6}$O$_{17}$ and KMo$_{6}$O$_{17}$.  Notably, all these four materials share a common structural motif, a zig-zag chain of corner-shared MoO$_6$ octahedra \cite {Canadell89}, along which a certain $t_{2g}$ orbital forms a 1d band, corresponding to the $b$ band for \MO.  In the case of the Li purple bronze, this is the only band dispersing up to the Fermi energy ($E_F$) and therefore the FS is almost perfectly 1d \cite {Denlinger99}.  However, for the other compounds, this band is one of the three bands that disperse to $E_F$, and therefore the FS is quasi-2d, albeit hidden-1d.  In the case of Na or K purple bronze, the other two bands are generated by three-fold rotations of the $b$ band, but, in the case of \MO, they ($b \pm c$ bands) are unrelated to the $b$ band by any crystal symmetry.  In fact, the $b \pm c$ bands have different fillings (about three-quarters filled) from that of the $b$ band (about half-filled, as in the other three materials).  In this paper, we will note some consequences of these similarities and differences for the CDW and NFL properties of \MO.

Angle-resolved photoemission spectroscopy (ARPES) is a valuable tool to directly measure the FS topology, as well as the size and the geometrical anisotropy of the gap function, as seen in other molybdenum oxides \cite{Gweon97} and high temperature superconductors \cite{Damascelli03}.  Also, by directly probing the single electron spectral function, the ARPES line shape provides detailed information about the nature of the single particle excitations of the system \cite{Denlinger99,Gweon01,Allen02,Gweon03}.  In past ARPES work on quasi-1d Li$_{0.9}$Mo$_{6}$O$_{17}$, and on hidden 1d NaMo$_{6}$O$_{17}$, we have directly observed their nested FS characters \cite {Gweon97, Denlinger99, Gweon01} and have identified certain fractionalization and NFL signatures \cite {Gweon03} in their ARPES line shapes.  

In this paper, we report the complete normal state FS geometry of \MO\ as measured by ARPES\@.  Based on the observed FS geometry, CDW nesting vectors \textbf{q}$_{c1}$ and \textbf{q}$_{c2}$ are explained by nesting of different parts of the normal state FS\@.  In particular, we find that, for explaining \textbf{q}$_{c2}$, the small 2d modulation of the $b$ band is a crucial factor.  We also report the associated spectra above $T_{c1}$ and the formation of the anisotropic gap function as temperature is decreased below $T_{c1}$, but not the changes that occur below $T_{c2}$, given our limited energy resolution.  Most important, we report and discuss NFL aspects of the ARPES line shapes, just like those found previously \cite {Gweon03} for hidden 1d Na purple bronze.  In particular, the near $E_F$ power law behavior for band $b$ is quite similar to that of the Na purple bronze, but is very different from that of the Li purple bronze.  This finding is consistent with the near perfect 1d FS for the Li purple bronze and the noticeable 2d modulations in the others.


Single crystalline samples of \MO\ were grown by a vapor transport technique \cite {Hill97}.  ARPES was performed at the 4m-NIM beamline of the Synchrotron Radiation Center (SRC) at the University of Wisconsin. Samples oriented by Laue diffraction were cleaved \textit{in situ} just before the measurement in a vacuum of $\approx$ 2.0 $\times$ 10$^{-10}$ Torr or better, exposing a clean surface of quasi-2d $bc$ plane. The integrity of the sample surface was continually monitored by the reproducibility of the spectrum. The sample temperature was controlled by a cryostat utilizing a closed-cycle He refrigerator with an embedded resistive heater. Monochromatized photons of energy 17 eV were used for the data reported in this paper.  Data taken in the photon energy range 8--20 eV, not shown, were consistent with a quasi-2d electronic structure expected for the layered structure, and so, throughout this paper, we will employ a 2d crystal approximation considering only a single tetragonal Mo$_6$O$_{22}$ layer.  In this approximation, $b \parallel b^*$ and $c \parallel c*$, the latter of which breaks down when the 3d monoclinic symmetry is taken into account.  The Fermi energy ($E_F$) and experimental energy resolution were calibrated by a reference spectrum taken on a freshly deposited Au thin film.  The total instrumental energy resolution \textit{$\Delta$E} was 40 meV and the full acceptance angle of the analyzer was 0.9$^{\circ}$, which amounts to 3 \% of the Brillouin zone dimension along the $b^*$ direction. 

The FS geometry was measured by making a momentum space $E_F$ intensity map. The experimental geometry was such that the sample orientation is fixed with respect to the direction of incidence of photons onto the sample and the electron energy analyzer was moved around the sample with angles well defined relative to the sample's crystal axes.  In the standard interpretation of ARPES, varying angles at a fixed kinetic energy gives access to momentum values of the photohole on a spherical shell in the momentum space.  Within our 2d crystal approximation, only the momentum values projected onto the $bc$ plane are meaningful.


\begin{figure}[!t]

\includegraphics[width=3.0 in]{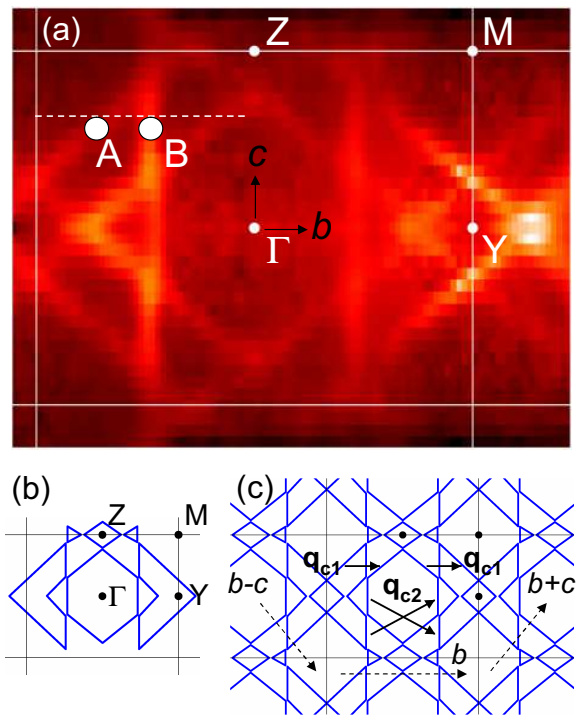}

\caption{(a) Fermi energy intensity map of \MO\ measured at T = 50 K using photon energy h$\nu$ = 17 eV.  (b) FS geometry extracted from the map of panel a, by tracing the intensity local maxima of the map.  Only one repeat unit is shown here.  (c) Same as (b), but repeated in many Brillouin zones.}

\end{figure}


In Figure 1, we address the normal state FS geometry.  Figure 1(a) shows the FS intensity map measured at T = 50 K\@.  High symmetry points of the tetragonal unit cell of the Mo$_6$O$_{22}$ layer are indicated.  The distance between $\Gamma$ and Y is 0.58 $\mathrm{\AA}^{-1}$ corresponding to $19^{\circ}$, and the distance between $\Gamma$ and Z is 0.47 $\mathrm{\AA}^{-1}$ corresponding to $15^{\circ}$.  The data were taken in $1^{\circ}$ steps in both angles for only the upper half of the whole FS map presented.  The full map was then generated by vertical reflection, a valid symmetry operation within the 2d crystal approximation.  In addition to this map, we have taken several maps at temperatures from below $T_{c2}$ to above $T_{c1}$.  However, given our energy resolution, all maps were readily interpreted as showing the normal state FS geometry, while the effect of the phase transition shows up only in a subtle change of intensity (see discussion of Figure 2 below).  

Figure 1(b) summarizes the FS geometry extracted from the map of Figure 1(a).  The best way to understand the FS is to use a repeated zone scheme, as in Figure 1(c).  To our knowledge, this is the first time that ARPES data have been presented to support a hidden-1d FS picture \cite{Canadell89,Hill97} for this material.  Indeed, the measured FS can be interpreted approximately as being composed of three 1d FSs defined by three 1d chain directions $b$ and $b \pm c$, marked by dashed arrows in Figure 1(c).  Also included in Figure 1(c) are wave vectors \textbf{q}$_{c1}$ and \textbf{q}$_{c2}$ observed by diffraction.  Given the observed hidden 1d FS, it is straightforward to deduce that the maximum nesting of FS would occur by a simultaneous nesting of $b \pm c$ bands with the wave vector $\approx$ \textbf{q}$_{c1}$, thus explaining the origin of the CDW transition at $T_{c1}$.  Next, for the 2nd CDW transition, we invoke a nesting scenario for the quasi-1d $b$ band.  As is well-known \cite {Gruner94}, when a 1d FS acquires a small periodic modulation perpendicular to the chain direction, maximum nesting condition is satisfied by connecting FS pieces at a well-defined angle relative to the 1d chain direction.  This is clearly the case in the current example, as the measured wave vector and its symmetry equivalent, the two arrows shown below the label \textbf{q}$_{c2}$, are approximately those wave vectors that maximally nest the FS arising from the $b$ band, given the slight zigzag of the FS\@.  Thus, our FS map nicely explains the origin of both CDW transitions.  Note that this explanation cannot yet give the shape of the FS geometry in either of the CDW phases.  To get at such details, data taken in each phase with far better momentum and energy resolutions would be necessary.


\begin{figure}[!t]

\includegraphics[width=3.0 in]{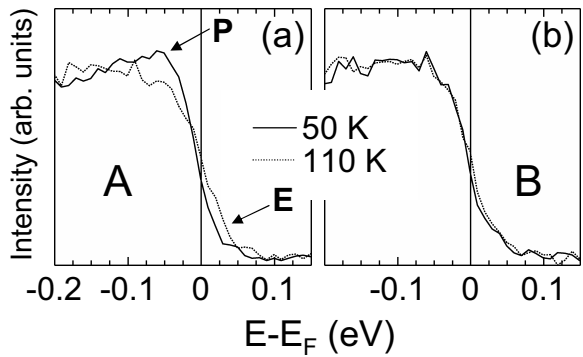}

\caption{Near $E_F$ spectra at two points on the FS, A and B of Figure 1(a).  Data are taken approximately at, and well below, $T_{c1}$ (109 K).  Leading edge shift (arrow ``E'') and intensity pile-up (arrow ``P'') occur at low temperature for point A (a), but not for point B (b).}

\end{figure}


Figure 2 shows the change of the valence band spectra as temperature is lowered below $T_{c1}$ for two points in \textbf{k}-space, labeled in Figure 1(a).  Point A is on the hidden-1d line which disappears below the CDW transition at $T_{c1}$, while point B is on a part of remnant FS that is not nested by \textbf{q$_{c1}$}.  As expected, and as observed previously \cite {Gweon98} with particular clarity in SmTe$_{3}$, little change is observed through the transition for point B where the nesting condition is not fulfilled, with the small observable change near $E_F$ consistent with a slight sharpening of the edge at low temperature.
At point A, we observe a leading edge shift, on the order of $\approx$ 15 meV, as well as intensity pile-up below $E_F$ over an energy scale of $\approx 50$ meV.  These spectral changes are qualitatively similar to what was observed in a quasi-1d CDW material K$_{0.3}$MoO$_3$ \cite {Gweon01}.  In the latter material, the spectral changes were observed over a much larger energy scale than that of the transition temperature scale.  Presumably due to the higher dimensionality, the extent of this scale difference is much reduced in the current material, by a factor of $\approx 4$ when the ratio of the intensity pile-up energy scale (50 meV) and the transition temperature scale (16 meV) \cite {Gap-scale} is compared to the equivalent ratio for K$_{0.3}$MoO$_3$.  Lastly, we point out that, despite the clearly observed CDW-induced changes in Figure 2(a), an energy resolution better than the current one (40 meV) would be necessary to address changes occurring in small energy scales.  For example, likely due to this finite energy resolution, the intensity at $E_F$ undergoes only a small reduction below $T_{c1}$ in the current data, which explains why the normal state FS geometry can be extracted from our FS map in Figure 1(a) taken well below $T_{c1}$.


\begin{figure}[!b]

\includegraphics[width=3.0 in]{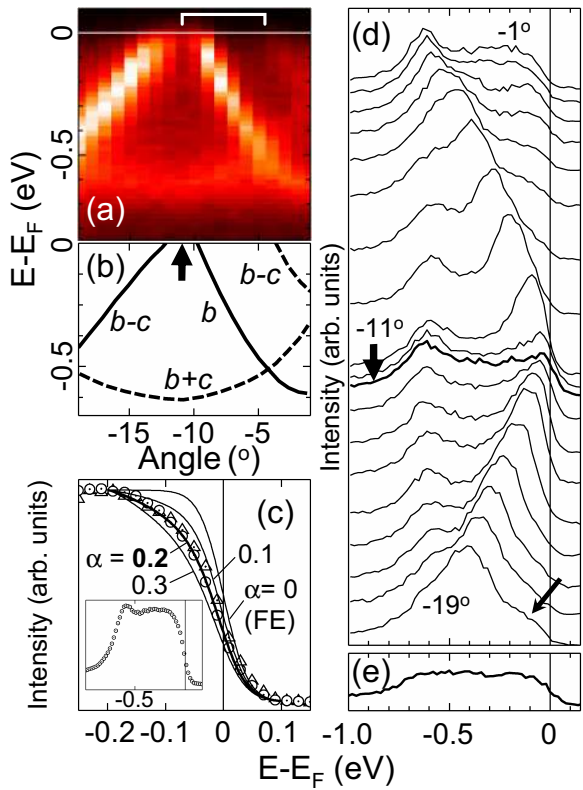}

\caption{(a) Angle-energy intensity map, where the angle range corresponds to momentum path along the dashed line shown in Figure 1(a).  (b) The dispersions of hidden 1d bands extracted from this map.  Each dispersion is labeled with the relevant 1d chain direction. For weaker features of the map, dispersions are given in dashed lines.  (c)  Angle-integrated spectrum of the map in (a), showing no Fermi edge (FE).  Data plotted in circles were obtained by angle integration over the entire angle range of (a), while data plotted in triangles were obtained by integration over a narrower range marked by the white bracket in (a), i.e.~they show near-$E_F$ behavior of the $b$ band alone.  Near-$E_F$ region is fit with a finite temperature Luttinger liquid theory line shape \cite {Orgad2001}, giving $\alpha$ (``anomalous dimension'') $\approx 0.2$.   The spectrum in the full energy window is shown as inset.  (d) Energy distribution curves (EDCs) for the map of (a).  Spectrum pointed by thick arrow corresponds to the angle pointed by thick arrow in (b).  (e) EDC at the Y point (see Figure 1(a)), where no occupied electron level exists according to the band calculation.  Data for (a,c,d) were taken at 200 K, while data for (e) were taken at 250 K\@.}

\end{figure}

Figure 3 shows detailed energy dispersions for the states contributing to the FS shown in Figure 1 and, more importantly, their anomalous NFL line shapes in the normal metallic phase well above $T_{c1}$.  The data correspond to the momentum space cut shown as a dashed line in Figure 1(a).  Consistent with the map of Figure 1(a), the data in Figure 3(a) show three FS crossings, as clearly indicated in Figure 3(b).  In addition, there is a slowly dispersing band well below $E_F$, dispersing along the 1d chain parallel to the $b+c$ direction and giving rise to the next $E_F$ crossings not covered by this cut but seen in Figure 1(a).  Thus this map verifies in detail the hidden 1d picture.

The NFL character of these two hidden 1d bands becomes apparent when the angle integrated spectrum is examined.  In Figure 3(c), the line shapes for the spectrum angle integrated over all bands (circles) or over the $b$ band alone (triangles) are compared with theoretical line shapes, calculated from the finite temperature Luttinger liquid (LL) spectral theory \cite {Orgad2001}, including full account of the instrumental energy resolution.  The comparison shows that the angle integrated spectrum in either case is best described by an LL line shape with a power law exponent 0.2, far from showing the Fermi edge expected for a Fermi liquid.  This exponent is similar to that \cite {Gweon03} of NaMo$_6$O$_{17}$, 0.3, but different from that \cite {Gweon03} of Li$_{0.9}$Mo$_{6}$O$_{17}$, 0.9, showing that the overall quasi-2d nature of the electronic structure has an important consequence on the properties of individual 1d bands.  

An additional possible NFL signature is the anomalous line shape {\em after} the band dispersion crosses $E_F$.  In Figure 3(d), the thick line highlights the line shape for angle -11$^\circ$, at which no intensity is expected from the band theory except for a peak at $\approx -0.6$ eV for the $b+c$ band.  Figure 3(e) shows a similar line shape with comparable intensity at the Y point (see Fig.~1(a)), where {\em no} occupied bands are expected in the band theory.  Because the Y point ($-11 ^\circ$ point) lies at the center of a $\pm 4^\circ$ ($\pm 1^\circ$; see Fig.~3(b)) window in which no bands are expected, the effect of the angle resolution window ($\pm 0.45 ^\circ$ or $\pm 0.55 ^\circ$ if the effect of the energy resolution 40 meV is also included) cannot explain the observed intensity. This line shape also cannot be due to inelastic photoelectron scattering, since its weight is confined to the bandwidth of the main bands.  Instead, line shape features observed at other momentum values (e.g.~see the feature marked by a slanted arrow in Fig.~3(d)) imply that this same line shape contributes additively to the spectra at all momentum values.  A similar feature has been seen in NaMo$_6$O$_{17}$ and a high $T_c$ cuprate, as discussed in Ref.~\onlinecite {Gweon03}, and more recently also in graphite \cite {Zhou05}.  The origin of this spectral feature is unknown, and possible explanations include elastic scattering by disorder/defect and quasi-elastic scattering by phonons, in both the initial and the final states of the photoemission process.  In the case of NaMo$_6$O$_{17}$ and the cuprate, we also offered \cite {Gweon03} a speculative interpretation in terms of electron fractionalization (``melted holon'') in the initial state, which is equally applicable to \MO.  While the validity of this last interpretation remains to be investigated further, we note that it is the only scenario that connects with the NFL line shape of the angle integrated spectrum.  In any case, a recent STM study \cite {Mallet01} shows defects and disorder on the cleaved surface of \MO, so the presence of disorder is another feature in common with the other materials.

In conclusion, the electronic structure of \MO\ as determined by ARPES displays the same properties that we have found for other hidden 1d molybdenum oxides, a hidden 1d FS that is consistent with band theory and with observed CDW transitions, and anomalous ARPES line shapes that yield no Fermi edge.

[{\em Erratum submitted, after the above text appeared in Phys.~Rev.~B.}]  We very much regret that in the above text we failed to cite angle resolved photoemission spectroscopy of \MO\ by another group \cite {takahashi, fujisawa, sato}.  In this work, as in ours, nesting and charge density wave (CDW) gap openings on different Fermi surfaces(FSs) were studied and conclusions very similar to ours were drawn about the CDW mechanism.  Some differences between our data and theirs can be noted.  (1) Better energy resolution in their data enabled detection of a small $b$ band gap opening.  (2) Sharper Fermi surface patterns in our data enabled the observation of small zigzags due to avoided FS crossings of the $b$ band, and hence a more detailed explanation of the nesting vector {\bf q}$_{c2}$. (3) For unknown reasons, the``P'' feature in our data (Fig.~2a), i.e. the intensity pile-up at low temperature, is not present in their data.  


\begin{acknowledgments}

This work was supported by the U.S. NSF at the University of
Michigan (UM) (Grant No.~DMR-03-02825), and at the SRC (Grant No.~DMR-00-84402).
\end{acknowledgments}

$^*$ Email address: gweon@umich.edu; Current address: Lawrence Berkeley National Laboratory, MS 2-200, 1 Cyclotron Road, Berkeley, CA 94720, USA.

$^\$$ Current address: Institute de Physique des Nanostructures, Ecole Polytechnique Fédérale de Lausanne, 1015 Lausanne, Switzerland

$^\dagger$ Current address: Los Alamos National Laboratory, Los Alamos, NM 87545, USA.

$^\ddagger$ Current address: Department of Physics, University of California Davis, Davis, California 95616, USA.


\begin{thebibliography}{99}
\bibitem{Gap-scale} This is the value of the photoemission gap $\Delta$, obtained from the standard weak-couping mean-field relation $\Delta = 1.76 k_B T_c$.
\end{thebibliography}
\end{document}